\documentstyle[preprint,aps]{revtex} 
 
\def\Journal#1#2#3#4{{#1} {\bf #2}, #3 (#4)} 
 
\def\NPB{{\em Nucl. Phys.} B} 
\def\PLB{{\em Phys. Lett.}  B} 
 
\def\PRD{{\em Phys. Rev.} D}

\begin{document} 
\draft
\preprint{
\parbox{4cm}{
\baselineskip=12pt
TMUP-HEL-9703\\ 
February, 1997\\
\hspace*{1cm}\\
\hspace*{1cm}
}}
\title{AN ANALYSIS ON A SUPERSYMMETRIC CHIRAL GAUGE THEORY
       WITH NO FLAT DIRECTION
\thanks{
Talk given at the International Workshop on Perspectives of
Strong Coupling Gauge Theories (SCGT96), November 1996, Nagoya, Japan
}}
\author{ N. KITAZAWA \thanks{e-mail: kitazawa@phys.metro-u.ac.jp}}
\address{Department of Physics, Tokyo Metropolitan University,\\
         Hachioji-shi, Tokyo 192-03, Japan}
\maketitle
\begin{abstract}
A guiding principle
 to determine the K\"ahler potential in the low energy effective theory
 of the supersymmetric chiral gauge theory with no flat direction
 is proposed.
The guiding principle is applied to
 the $SU(5)$ gauge theory with chiral superfields
 in the $5^*$ and $10$ representation,
 and the low energy effective theory is consistently constructed.
The spontaneous supersymmetry breaking
 takes place in the low energy effective theory as expected.
The particle mass spectrum in the low energy is explicitly calculated.
\end{abstract} 
\newpage
 
\section{Introduction} 

It is the long-standing subject
 to systematically evaluate the non-perturbative effect of the gauge theory
 in four dimension.
In the non-supersymmetric theory
 we have to use some non-systematic truncation,
 and only the qualitative feature of the non-perturbative effect is known.
On the other hand,
 a breakthrough occurred in the method of the analysis
 of the supersymmetric gauge theory.
For example,
 the low energy effective theory
 of the $N=2$ supersymmetric $SU(2)$ Yang-Mills theory
 is exactly determined~\cite{SW},
 and the superpotential of the low energy effective theory
 of the $N=1$ supersymmetric QCD
 is also exactly determined~\cite{S1,S2,ILS}.
The method is also applied to understand the interesting phenomena
 of the dynamical supersymmetry breaking
 by the strong gauge interaction,~\cite{ISS,PT1,IY1,IT,PST}
 and the result is used to construct
 some concrete models.~\cite{DNS,IY2,PT2,KO}

The supersymmetric chiral gauge theory with no flat direction,
 especially the $SU(5)$ gauge theory with the chiral superfields
 in the $5^*$ and $10$ representation,
 has been extensively considered as the system
 where the dynamical supersymmetry breaking can be expected.
Although the new method is powerful to obtain the low energy effective theory
 of the $SU(5)$ gauge theory with two pairs of the chiral superfields
 in the $5^*$ and $10$ representation,~\cite{M,V}
 it is not straightforward to apply to the $SU(5)$ gauge theory
 with one pair of the chiral superfields in the $5^*$ and $10$ representation,
 because the effect of the non-trivial K\"ahler potential
 is important at low energy.~\cite{ADS}
The new method,
 the principle of the symmetry and holomorphy,
 is powerful to determine the superpotential,
 but it does not give any constraint to the K\"ahler potential.

In this paper we present a guiding principle to determine the
 effective K\"ahler potential of the low energy effective theory
 of the supersymmetric chiral gauge theory with no flat direction.~\cite{K}
As an example,
 the low energy effective theory of the the $SU(5)$ gauge theory
 with the one pair of the chiral superfields
 in the $5^*$ and $10$ representations is constructed.
 
\section{Fundamentals of the theory and low energy effective fields}

We first summarize the classical properties
 of the supersymmetric $SU(5)$ gauge theory
 with chiral superfields in the $5^*$ and $10$ representations.

The anomaly-free global symmetry of this system is $U(1)_R \times U(1)_A$.
The charges of the fields are as follows.
\begin{equation}
\begin{array}{ccc}            
                            & ~U(1)_R~~   & ~U(1)_A~~ \\
 ~~~W^{\dot\alpha}~~~       &    -1       &    0      \\
 ~~~\Phi~~~                 &    9        &    3      \\
 ~~~\Omega~~~               &    -1       &    -1     \\
\end{array}
\end{equation}
Here $W^{\dot\alpha}$
 is the chiral superfield of the $SU(5)$ gauge field strength,
 and $\Phi$ and $\Omega$ denote the chiral superfields of the matter
 in the $5^*$ and $10$ representations, respectively.
The classical scaler potential
 comes from the D-component of the vector superfields,
\begin{equation}
 V_D = {1 \over {2g^2}} D^a D^a
\label{classical-in-original}
\end{equation}
 with
\begin{eqnarray}
 D^a = g^2 \left\{ A_\Phi^{\dag} T_{5^*}^a A_\Phi
                 + A_\Omega^{\dag} T_{10}^a A_\Omega
           \right\},
\end{eqnarray}
 where $g$ denotes the gauge coupling constant,
 and $A_\Phi$ and $A_\Omega$ are the scalar components
 of the chiral superfields $\Phi$ and $\Omega$,
 and $T_{5^*}^a$ and $T_{10}^a$ denote the generators of $SU(5)$
 for the $5^*$ and $10$ representations, respectively.
It is well known that
 $A_\Phi=0$ and $A_\Omega=0$ is the unique solution
 of the stationary condition of this potential,
 and the classical vacuum is supersymmetric.

It is remarkable that
 no gauge invariant superpotential can be written down.
Since all the gauge invariant holomorphic polynomial
 composed by the chiral superfields $\Phi$ and $\Omega$ vanish,
 we can not consider non-trivial superpotential
 even the non-renormalizable one.
This fact means that the gauge coupling $g$
 or the scale of the dynamics $\Lambda$ is a unique parameter
 in this theory.

Now we consider what effective fields are appropriate in this theory.
Since we do not know the symmetry of the exact vacuum
 unlike in the case of supersymmetric QCD,
 we must assume it.
Here we assume that
 both $U(1)_R$ and $U(1)_A$ symmetry are spontaneously broken,
 and there is no massless fermions except for the Nambu-Goldstone fermion
 due to the spontaneous breaking of supersymmetry.
Therefore, 't Hooft anomaly matching condition is not imposed.
Furthermore,
 the confinement of $SU(4)$, a subgroup of $SU(5)$,
 is assumed rather than the confinement of $SU(5)$ itself.
This is the assumption of the complementarity,
 namely, we will proceed the analysis in Higgs phase
 rather than the confining phase.
These assumptions have to be justified by the result of the analysis.

We introduce only the effective fields
 which couple with the Lorentz invariant bi-linear operators
 composed by the three fields $\Phi$, $\Omega$ and $W$ in the original theory.
In addition,
 we assume that the effective fields are
 in the smallest representations of $SU(5)$ in each bi-linear combinations.
Namely, we consider the following three effective fields.
\begin{equation}
 \begin{array}{lll}
  X \equiv X^{i=5} \quad & X^i \sim \epsilon^{ijklm} \Omega_{jk} \Omega_{lm} &
  \qquad 10 \times 10 \rightarrow 5^* \\
  Y \equiv Y_{i=5} \quad & Y_i \sim \Phi^j \Omega_{ji} &
  \qquad 5^* \times 10 \rightarrow 5 \\
  S & S \sim {\rm tr} \left( W^{\dot\alpha} W_{\dot\alpha} \right) &
  \qquad 24 \times 24 \rightarrow 1
 \end{array}
\end{equation}
The operator corresponding $5^* \times 5^* \rightarrow 10^*$ vanishes,
 since the superfields commute each other.
Since we assume the confinement of $SU(4)$,
 only the $SU(4)$-singlet parts of each effective fields
 are introduced as the effective fields.

This procedure is supported by the following arguments.
The classical scalar potential eq.(\ref{classical-in-original})
 can be written like
\begin{eqnarray}
 V_D &=& {{g^2} \over 2}
       \left[
        \left( A_\Omega^{\dag} T^a_{10} A_\Omega \right)
        \left( A_\Omega^{\dag} T^a_{10} A_\Omega \right) + \cdots
       \right]
\\
     &=& {{g^2} \over 2}
       \left[ -\lambda(10,10,5^*){1 \over {4!}}
               \left| \epsilon A_\Omega A_\Omega \right|^2
              -\lambda(10,10,50)
               \left| \left( A_\Omega A_\Omega \right)_{50} \right|^2
              + \cdots
       \right],
\nonumber
\end{eqnarray}
 where $\lambda(r_1,r_2,r_c) \equiv \{ C_2(r_1)+C_2(r_2)-C_2(r_c) \}/2$,
 and $C_2(r)$ denotes the coefficient of the second Casimir invariant
 of the representation $r$ of $SU(5)$.
 \footnote{The operator correspond to the channel
           $10 \times 10 \rightarrow 45$ vanishes
           because of the Bose statistics of the scalar field.}
The method of the auxiliary field
 can be used to introduce the effective fields.
\begin{eqnarray}
 V_D &\rightarrow& V_D
    + {1 \over 2}\lambda(10,10,5^*){1 \over {4!}}
      \left| \ g \ \epsilon A_\Omega A_\Omega - A_X \right|^2
    + \cdots
\nonumber\\
  &=& {1 \over 2}\lambda(10,10,5^*){1 \over {4!}} A_X^i A_{Xi}^{\dag}
    - {g \over 2}\lambda(10,10,5^*){1 \over {4!}}
      \left\{
        \big( \epsilon A_\Omega A_\Omega \big)^i A_{Xi}^{\dag}
       + {\rm h.c.}
      \right\}
\nonumber\\
  &&+ \cdots,
\end{eqnarray}
 where $A_X^i$ denotes the scalar component of the effective field $X^i$.
This result shows that
 if the coefficient $\lambda$ is positive,
 the classical squared mass of the effective field becomes positive,
 and it is worth considering.
The effective field in the $5^*$ representation can be considered,
 since $\lambda(10,10,5^*) = {{12} \over 5} > 0$,
 but the effective field in the $50$ representation can not be considered,
 since $\lambda(10,10,50) < 0$, and its classical squared mass is negative.
The same arguments are true
 for the effective fields composed by $\Phi$ and $\Omega$.
The effective field $Y_i$ is worth considering,
 since $\lambda(5^*,10,5) = {9 \over 5} >0$,
 but the effective field in the $45$ representation can not be considered,
 since $\lambda(5^*,10,45) < 0$.

From this argument
 we obtain the classical scalar potential
 written by the effective fields $A_X$ and $A_Y$.
\begin{equation}
 V_{\rm classical}
  = \lambda_X |A_X|^2 + \lambda_Y |A_Y|^2,
\label{classical-in-effective}
\end{equation}
 where $\lambda_X \equiv {1 \over 2} {1 \over {4!}} \lambda(10,10,5^*)$
 and $\lambda_Y \equiv {1 \over 5} \lambda(5^*,10,5)$.

\section{Superpotential and K\"ahler potential}

The general form of the superpotential
 is obtained by the requirement of the symmetry and holomorphy as
\begin{equation}
 W = S f \left( {{\Lambda^{13}} \over {XYS^3}} \right)
\end{equation}
 with a general holomorphic function $f$.
Note that the power of $\Lambda$, $13$,
 which comes from the dimensional analysis,
 is just the coefficient of the 1-loop $\beta$-function
 of the $SU(5)$ gauge coupling.
In the weak coupling limit, $\Lambda \rightarrow 0$,
 this superpotential must coincide with the gauge kinetic term
 in the perturbatively-calculated Wilsonian action of the original theory.
From this condition,
\begin{equation}
 W = - {1 \over {64\pi^2}}
       \ S \ \ln \left( {{\Lambda^{13}} \over {XYS^3}} \right)
     + S {\tilde f} \left( {{\Lambda^{13}} \over {XYS^3}} \right),
\end{equation}
 where $\tilde f$ is a holomorphic function
 with $\lim_{z \rightarrow 0} {\tilde f}(z) = 0$.
Moreover,
 since we are considering that
 the massless degrees of freedom are only the Nambu-Goldstone particles,
 and all of them are described by the effective fields already introduced,
 the function $\tilde f$ should not have the singularities,
 and it is a constant.
The constant is absorbed to the redefinition of $\Lambda$.
Thus, we obtain
\begin{equation}
 W = - {1 \over {64\pi^2}}
       \ S \ \ln \left( {{\Lambda^{13}} \over {XYS^3}} \right).
\label{superpotential}
\end{equation}
This is the unique superpotential within our postulations.

We propose the following two conditions
 to constrain the K\"ahler potential in the effective theory.
\begin{enumerate}
 \item The K\"ahler potential coincides with the naive one
       described by the effective fields
       in the limit of weak strength of the effective fields,
 \item The scalar potential coincides with the classical one
       in the limit of weak coupling.
\end{enumerate}
The first condition is necessary
 so that the effective fields are the quantizable fields
 with canonical kinetic terms.
The second condition requires the classical scaler potential
 which is described by the effective fields
 like the potential of eq.(\ref{classical-in-effective}).
Here, we demonstrate constructing the non-trivial K\"ahler potential
 within the ansatz of factorization:
\begin{equation}
 K(X^{\dag}X,Y^{\dag}Y,S^{\dag}S)
  = K_X(X^{\dag}X) + K_Y(Y^{\dag}Y) + K_S(S^{\dag}S).
\label{kaehler}
\end{equation}

We can consider the following K\"ahler potential
 which satisfies the above two conditions.
\begin{eqnarray}
&
 K_X(X^{\dag}X) = {1 \over {\Lambda^2}} f(X^{\dag}X)_{C_X/\Lambda^4},
\quad
 K_Y(Y^{\dag}Y) = {1 \over {\Lambda^2}} f(Y^{\dag}Y)_{C_Y/\Lambda^4},
&
\\
&
 K_S(S^{\dag}S) = {1 \over {\Lambda^4}} S^{\dag}S,
&
\end{eqnarray}
 where $C_X$ and $C_Y$ are real parameters,
 and a function $f(z)_a$ is defined by
\begin{equation}
 f(z)_a \equiv \sum_{n=0}^\infty (-1)^n {{a^{2n}z^{2n+1}} \over {(2n+1)^2}}
        = z \ F \left(1, {1 \over 2}, {1 \over 2};
                      {3 \over 2}, {3 \over 2}; -a^2 z^2 \right).
\end{equation}
The function $F$ is the generalized hypergeometric function.
Note that
 the K\"ahler potentials $K_X$ and $K_Y$ become naive ones
 in the weak field limit of $C_X X^{\dag}X / \Lambda^4 \rightarrow 0$
 and $C_Y Y^{\dag}Y / \Lambda^4 \rightarrow 0$, respectively
 (first condition).
 \footnote{
  This limit can also be understood
   as the limit of $\Lambda \rightarrow \infty$.
  If this limit can be regarded as the strong coupling limit,
   the condition of the coincidence with the naive one may seem to be strange.
  However,
   it is not clear whether this limit is really the strong coupling limit,
   since we do not know the value of the effective coupling
   below the scale of $\Lambda$.
  In addition,
   note that ``naive'' does not mean ``tree'',
   namely no quantum correction.
  The naive K\"ahler potential
   is described by the composite effective fields.}
The scalar potential is obtained as
\begin{eqnarray}
 V &=& {{\Lambda^4} \over {(64 \pi^2)^2}}
       \left|
        \ln \left( {{A_X A_Y A_S^3} \over {\Lambda^{13}}} \right) + 3
       \right|^2
    +  {{\Lambda^2} \over {(64 \pi^2)^2}}
       \left(
        {{|A_S|^2} \over {|A_X|^2}} + {{|A_S|^2} \over {|A_Y|^2}}
       \right)
\nonumber\\
   &&+ {{C_X^2} \over {(64 \pi^2)^2}} {{|A_S|^2} \over {\Lambda^6}} |A_X|^2
     + {{C_Y^2} \over {(64 \pi^2)^2}} {{|A_S|^2} \over {\Lambda^6}} |A_Y|^2.
\label{potential}
\end{eqnarray}
The last two terms are the contribution of the non-trivial K\"ahler potential.

The two parameters $C_X$ and $C_Y$ are determined
 so that the potential of eq.(\ref{potential})
 coincides with the classical one, eq.(\ref{classical-in-effective}),
 in $\Lambda \rightarrow 0$ limit (second condition).
The first two terms of the potential simply vanish in this limit,
 but the last two terms seem to be singular.
We integrate out the effective field $S$
 by replacing the field by its vacuum expectation value.
The vacuum expectation value of $S$ is proportional to $\Lambda^3$,
 and the coefficient $r$ is independent of $\Lambda$,
 but it depends on $C_X$ and $C_Y$.
Therefore,
 we can determine these two parameters by the conditions
\begin{equation}
 {{C_X^2} \over {(64 \pi^2)^2}} \ r(C_X,C_Y)^2 = \lambda_X,
\quad
 {{C_Y^2} \over {(64 \pi^2)^2}} \ r(C_X,C_Y)^2 = \lambda_Y,
\end{equation}
 which are obtained by the second condition.
The construction of the effective theory is finished.

\section{Vacuum and mass spectrum}

The vacuum expectation values of the effective fields
 are obtained by solving the stationary conditions of the potential
 of eq.(\ref{potential}).
It is analytically shown that
 all the vacuum expectation values are real and positive,
 and the numerical calculation gives
\begin{equation}
 \langle A_X \rangle \simeq (0.17)^2,
\quad
 \langle A_Y \rangle \simeq (0.11)^2,
\quad
 \langle A_S \rangle \simeq (0.31)^3,
\label{solution}
\end{equation}
 in unit of $\Lambda$.
This solution
 is consistent with the assumption of breaking $SU(5) \rightarrow SU(4)$,
 since the effective fields $X$ and $Y$,
 which are the components of the effective field
 in the $5^*$ and $5$ representations of $SU(5)$, respectively,
 obtain the vacuum expectation values.
The assumption of the complete breaking 
 of the global $U(1)_R \times U(1)_A$ symmetry is also confirmed.
Since the vacuum expectation value of the effective filed $S$
 means the gaugino pair condensation,
 the spontaneous breaking of supersymmetry is expected
 through Konishi anomaly.~\cite{konishi}
In fact,
 the vacuum energy density is not zero,
 $V_{\rm vacuum} \simeq (0.16)^4$ in unit of $\Lambda$.

The mass spectrum of the effective fields can be explicitly calculated.

On boson fields,
 it is convenient to consider the non-linear realization
 of the global $U(1)_R \times U(1)_A$ symmetry:
\begin{equation}
 A_X = \Lambda \ \phi_X e^{i\theta_X/\Lambda},
\quad
 A_Y = \Lambda \ \phi_Y e^{i\theta_Y/\Lambda},
\quad
 A_S = \Lambda^2 \ \phi_S e^{i\theta_S/\Lambda},
\end{equation}
 where $\phi_{X,Y,S}$ and $\theta_{X,Y,S}$ are the real scalar fields
 with dimension one.
The eigenvalues of the mass matrix for the real scalar fields
 $\theta_{X,Y,S}$ are analytically obtained.
Two of three eigenvalues are zero
 which are corresponding to the Nambu-Goldstone bosons
 of $U(1)_R$ and $U(1)_A$ breaking,
 and remaining eigenvalue is
 $m_\theta^2 = 22\Lambda^2/(64 \pi^2)^2 \simeq (0.0074 \Lambda)^2$.
The smallness of this value can be understood by considering that
 it is corresponding to the mass of the pseudo-Nambu-Goldstone boson
 due to the anomalous global $U(1)$ symmetry breaking.
The eigenvalues of the mass matrix for the fields $\phi_{X,Y,S}$
 are numerically obtained as
\begin{equation}
 m_\phi^2 \simeq (0.45)^2, \quad (0.73)^2, \quad (1.5)^2,
\end{equation}
 in unit of $\Lambda$.

The mass matrix of the fermion components,
 $\Lambda \psi_X$, $\Lambda \psi_Y$ and $\Lambda^2 \psi_S$,
 of the effective chiral superfields, $X$, $Y$ and $S$, respectively,
 is analytically obtained, where all $\psi$'s have dimension $3/2$.
One can analytically check that
 the mass matrix has one zero eigenvalue,
 which is corresponding to the mass of the Nambu-Goldstone fermion
 of the supersymmetry breaking,
 by using the stationary conditions of the scalar potential.
The other two eigenvalues are numerically given by
\begin{equation}
 m_\psi \simeq 0.33, \quad 0.091,
\end{equation}
 in unit of $\Lambda$.

\end{document}